\documentclass[conference]{IEEEtran}
\IEEEoverridecommandlockouts
\usepackage{cite}
\usepackage[left=1.72cm,right=1.72cm,top=0.71in,bottom=0.96in]{geometry}
\usepackage{amsmath,amssymb,amsfonts,mathtools,bbm}
\usepackage{physics}

\usepackage{algorithmic}
\usepackage{graphicx}
\usepackage{textcomp}
\usepackage{xcolor}
\usepackage{adjustbox}
\usepackage{graphicx}

\usepackage{tikz}
\tikzset{>=latex}

\usetikzlibrary{
    quantikz,
    shapes, 
    arrows,
    positioning
}

\usepackage{pgf}
\usepackage{pgfplots}
\pgfplotsset{compat=1.3}   
\usepgfplotslibrary{colorbrewer}

\PassOptionsToPackage{hyphens}{url}
\usepackage[bookmarks=false]{hyperref}
\usepackage{cite}
\usepackage[capitalize]{cleveref}

\usepackage{xcolor}
\usepackage[
    shortcuts,       % enable short forms like gls etc.
    acronym,         % enable auto-generation of the acronym glossary
    nomain,          % disable the main glossary
    section=chapter, % set the glossary title format to be like a chapter (alternatives are section, subsection, etc.)
    toc=true,        % show in table of contents
]{glossaries}
\glsenableentrycount % enable counting of the references, otherwise \cgls does the same as \gls

\definecolor{mypurple}{rgb}{0.49,0.18,0.56}
\definecolor{mygold}{rgb}{0.93,0.49,0.13}
\definecolor{mygreen}{rgb}{0,0.5,0}
\definecolor{myblue}{rgb}{0,0,0.75}
\definecolor{mymagenta}{cmyk}{0,1,0,0.12}
\definecolor{mygray}{rgb}{0.5,0.5,0.5}

\begin{document}

\title{Testing of Hybrid Quantum-Classical K-Means for Nonlinear Noise Mitigation}

\author{%
    \IEEEauthorblockN{%
    Ark Modi\IEEEauthorrefmark{1},
    Alonso Viladomat Jasso\IEEEauthorrefmark{2},
    Roberto Ferrara\IEEEauthorrefmark{1},
    Christian Deppe\IEEEauthorrefmark{1},
    Janis Nötzel\IEEEauthorrefmark{2},\\
    Fred Fung\IEEEauthorrefmark{3},
    Maximilian Schädler\IEEEauthorrefmark{3}
    }
    \IEEEauthorblockA{%
    \IEEEauthorrefmark{1}
    Institute for Communications Engineering (LNT),
    \\
    \IEEEauthorrefmark{2}
    Emmy Noether Group for Theoretical Quantum Systems Design
    \\
    Technical University of Munich, D-80333 Munich, Germany
    \\
    \IEEEauthorrefmark{3}
    Optical and Quantum Laboratory, Munich Research Center\\
    Huawei Technologies D\"usseldorf GmbH, Riesstr. 25-C3,80992 Munich, Germany
    \\
    Email:
    \{ark.modi, viladomat.jasso, roberto.ferrara, christian.deppe, janis.noetzel\}@tum.de
    \\
    \{fred.fung, maximilian.schaedler\}@huawei.com}
}

\maketitle

\begin{abstract}
Nearest-neighbour clustering is a powerful set of heuristic algorithms that find natural application in the decoding of signals transmitted using the $M$-Quadrature Amplitude Modulation ($M$-QAM) protocol.
Lloyd et al.\ proposed a quantum version of the algorithm that promised an exponential speed-up. 
We analyse the performance of this algorithm by simulating the use of a hybrid quantum-classical implementation of it upon 16-QAM and experimental 64-QAM data.
We then benchmark the implementation against the classical k-means clustering algorithm.
The choice of quantum encoding of the classical data plays a significant role in the performance, as it would for the hybrid quantum-classical implementation of any quantum machine learning algorithm.   
In this work, we use the popular angle embedding method for data embedding and the swap test for overlap estimation. 
The algorithm is emulated in software using Qiskit and tested on simulated and real-world experimental data. 
The discrepancy in accuracy from the perspective of the induced metric of the angle embedding method is discussed, and a thorough analysis regarding the angle embedding method in the context of distance estimation is provided. 
We detail an experimental optic fibre setup as well, from which we collect 64-QAM data. 
This is the dataset upon which the algorithms are benchmarked.
Finally, some promising current and future directions for further research are discussed.
\end{abstract}

\begin{IEEEkeywords}
Quantum k nearest-neighbour,
Quantum Machine Learning, 
Quantum Computing, 
$k$-Means Clustering,
6G Communication,
Quadrature Amplitude Modulation, 
Quantum-Classical Hybrid Algorithms
\end{IEEEkeywords}

\section{Introduction}

    Quantum information processing is a method that started out to revolutionize the current theory of computation. Much has been done in showing its theoretical potential, with several candidate algorithms offering quadratic, exponential or even greater speed-ups. One question is whether part of this potential can be exploited with NISQ devices, which is often been impeded by hurdles during implementation. Hybrid-quantum classical systems are a popular option for the implementation of quantum algorithms due to the lack of stable quantum memory (QRAM) and due to the realisation that only certain tasks are suitable to be offloaded for quantum processing. Here, we compare the classical and hybrid quantum-classical implementation of a machine learning algorithm applied to the decoding problem of classical optical-fiber communication. 
    
    Quantum-enhanced Machine Learning, using quantum algorithms to learn quantum or classical systems, has sometimes promised even exponential speed-ups over classical machine learning. 
    Due to the significance of classical machine learning today, there is a lot of focus on such quantum machine learning algorithms for tomorrow's world. 
    A number of works such as ~\cite{googleSupremacy,SchuldPetruccione2018, schuld2014, kerenidis2017recommendation, kerenidis2018q, lloyd2013quantum} have showcased theoretical algorithms and experimental implementations that naturally give confidence that other quantum algorithms, including quantum machine learning ones, might eventually lead to industrial quantum algorithms with a speed up. 
    Many of these methods claim to offer exponential speedups over the analogous classical algorithms. 
    However there exist significant gaps between theoretical prediction and implementation.
    
    We demonstrate the problems and possible opportunities when applying the quantum $k$ nearest-neighbour clustering algorithm to the problem of decoding M-QAM signals.  
    It is known that the $k$-means clustering algorithm can be used for phase estimation in optical fibers~\cite{pakala2015non,zhang2017k}.
    Though the quantum version of this algorithm~\cite{lloyd2013quantum} promises an exponential speed-up, its usefulness in NISQ devices and end-to-end classical systems is under debate~\cite{Tang_PCA_2021, kerenidis2018q}.
    Due to such difficulties, we study the advantages and drawbacks as seen from a the M-QAM application.

    \tikzset{%
	block/.style = {draw,rounded corners, fill=white, rectangle, minimum height=2em, minimum width=6em},
	EDFA/.style = {draw, fill=white, regular polygon, regular polygon sides=3,minimum size=1.1cm},
	sum/.style= {draw, fill=none, circle, node distance=0.5cm,color=black,minimum size=14pt},
	fiber/.style= {draw, fill=none, circle, node distance=2cm,color=red,minimum size=20pt},
}

\begin{figure*}
	\centering
	\footnotesize
    
	\begin{center}
    % \begin{adjustbox}{max width = 0.5\textwidth}
		\begin{tikzpicture}[ >=latex']

			%\node [block, right=0.25cm of Mapper, align = center] (ECL1) {ECL};
			\node [block, align = center,minimum width=4em]  at (0,0) (ECL1) {ECL};
			\node [block,right=0.17cm of ECL1, align = center] (IQM) {DP-IQM};
			\node [block,above=0.17cm of IQM, align = center] (driver) {Drivers};
			\node [block,above=0.17cm of driver, align = center] (AWG) {AWG};
			\node [block,dashed,left=0.17cm of AWG, align = center] (TXDSP) {Tx-DSP};
			\node [EDFA, right=0.2cm of IQM,align = center,rotate=0,shape border rotate=-90] (EDFA_TX) {\kern-0.5emEDFA\kern-1.0em};
			
			\node [fiber,align = center] at (4.5,0.23) (fiber1) {};	
			\node [fiber, right=-0.6cm of fiber1,align = center] (fiber2) {};
			\node [fiber, right=-0.6cm of fiber2,align = center] (fiber3) {};
			\node [label, above = 0.1cm of fiber1, draw=none,align = center] (label1) {\tiny 4$ \times$ 20km\\ \tiny G.652 };	
			\node [EDFA, right=2.0cm of EDFA_TX,align = center,rotate=0,shape border rotate=-90] (EDFA_RX) {\kern-0.5emEDFA\kern-1.0em};

			\node [block,right=0.4cm of EDFA_RX, align = center] (HB) {$90^\circ$ Hybrid};
			\node [block,above=0.17cm of HB, align = center] (PD) {Photodiodes};
			\node [block,right=0.17cm of HB, align = center] (ECL2) {ECL};
			\node [block,above=0.17cm of PD, align = center] (Scope) {Oscilloscope};
			\node [block,dashed,right=0.17cm of Scope, align = center] (RXDSP) {CD $\mapsto$ CFO $\mapsto$ MIMO $\mapsto$ TR$\&$CPE};
			\node [label, below left = 0.0cm and -0.9cm of RXDSP, draw=none] (label1) {\tiny Rx-DSP};
			\node [label, right=0.25cm of RXDSP, align = center] (label3) {Dataset};			
			\draw [-latex,color=black] (RXDSP) -- (label3);
			
			\draw [-latex,color=red] (ECL1) -- (IQM);
			\draw [-latex,color=red] (IQM) -- (EDFA_TX);
			\draw [-latex,color=red] (EDFA_TX) -- (EDFA_RX);
			\draw [-latex,color=red] (EDFA_RX) -- (HB);
			\draw [-latex,color=red] (ECL2) -- (HB);

			\draw [-latex,color=black] (TXDSP) -- (AWG);
			\draw [-latex,color=black] (Scope) -- (RXDSP);
			
			\foreach \j in  {0.3,0.1,-0.1,-0.3}
			\draw [-latex, color=red] (HB.north) + (\j,0) -- ++(\j,0.12);
			\foreach \j in  {0.3,0.1,-0.1,-0.3}
			\draw [-latex, color=black] (PD.north) + (\j,0) -- ++(\j,0.12);

			\foreach \j in  {0.24,0.08,-0.08,-0.24}
			\draw [-latex, color=black] (AWG.south) + (\j,0) -- ++(\j,-0.12);
			
			\foreach \j in  {0.24,0.08,-0.08,-0.24}
			\draw [-latex, color=black] (driver.south) + (\j,0) -- ++(\j,-0.12);
			
		\end{tikzpicture}
    %   \end{adjustbox}
	\end{center}

	\caption{Experimental setup over a 80 km G.652 fiber link at optimal launch power of 6.6 dBm.} 
	\label{fig:Setup}
\end{figure*}
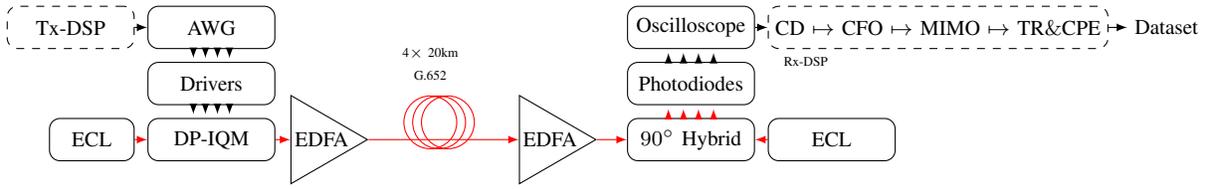

    In \cref{sec:problem}, we review the problem of signal processing through optic fibre cables and detail the experimental setup. % used to collect the 64-QAM data and generate the 16-QAM data.
    In \cref{sec:methods}, we describe the encoding method and the quantum circuit used for simulation.
    After this, the results of the simulation of hybrid quantum-classical $k$-means %acting upon both 16-QAM and experimental 64-QAM data vis-\'a-vis classical k-means clsutering 
    are described in \cref{sec:results}.
    We end with some promising current and future directions for further research~\cite{Tang_PCA_2021, fanizzaRosati-beyond-swap, SchurSamplingHarrow, efficientQCforSchurTxandCGT, QuantumSchurTransform2007, HighDimSchurTx, stephen2021, stereo_paper_entropy}. 
    
\section{Problem and Data Description}
    \label{sec:problem}
    
    In QAM multiple bits are conveyed in each time interval and carrier symbol by dividing the phase space of the carrier wave of fixed frequency, and designating each space to a unique bitstring.
    Signals are prepared to lie in the phase space corresponding to the required bitstring by modulating the amplitude of two carrier waves separated by 90$^\circ$ (for example, sine and cosine waves of the same frequency) and superposing them. 
    The receiver coherently separates the waves by using the orthogonality of the two waves. 
    QAM can achieve arbitrarily high spectral efficiencies by setting a suitable constellation size, limited only by the noise level and linearity of the channel~\cite{QAMbarnard}.
    
\subsection{Experimental Setup for Data Collection} 
    
\begin{figure}
    \begin{minipage}[H]{0.48\textwidth}
    \includegraphics[width=\textwidth]{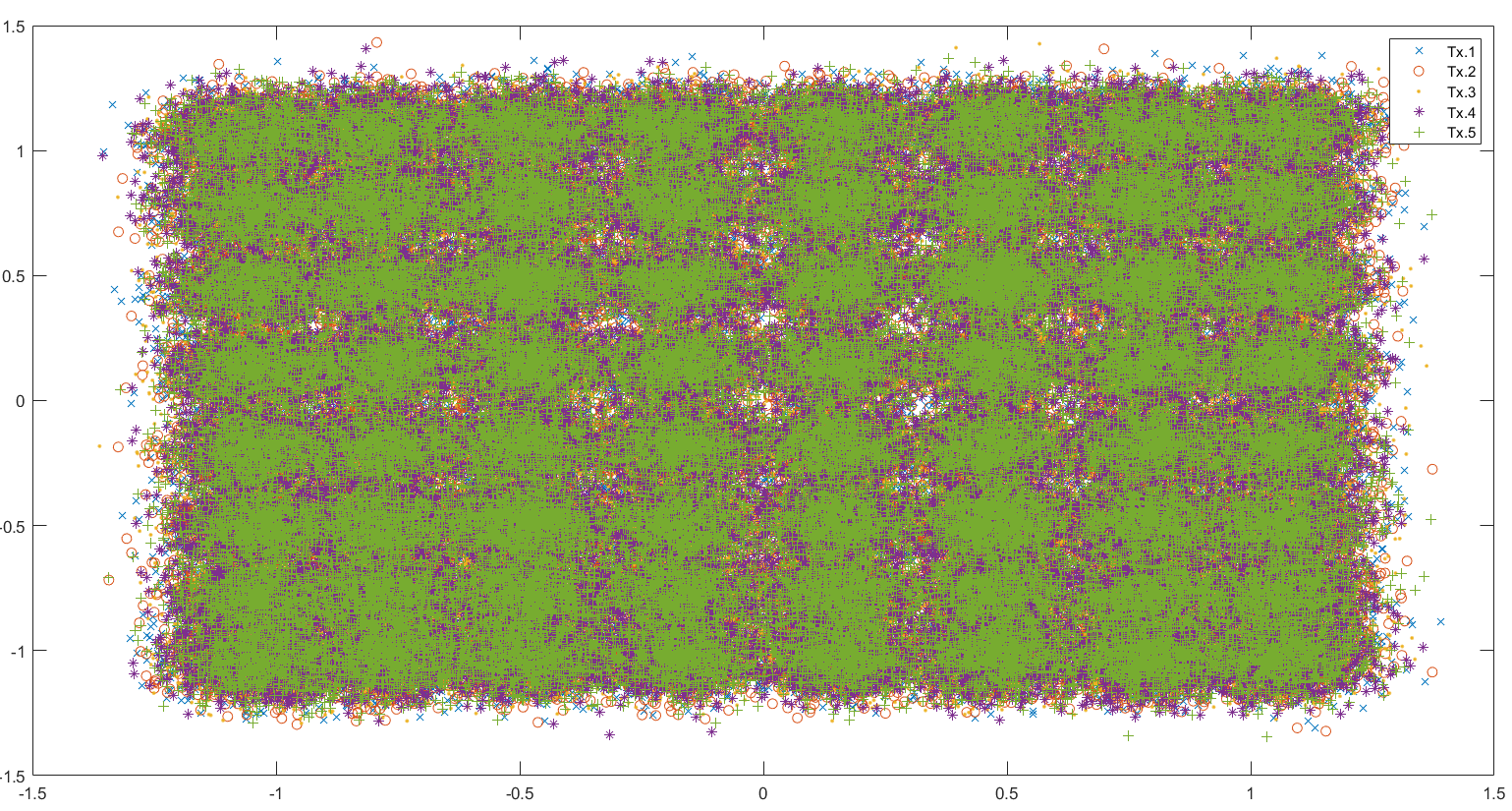}
        \caption{The data detected by the receiver from the least noisy (2.7dBm noise) channel. All 5 iterations of transmission are depicted together.}
        \label{fig:chap_data_2_7}
    \end{minipage}\\%
    \begin{minipage}[H]{0.48\textwidth}
    \includegraphics[width=\textwidth]{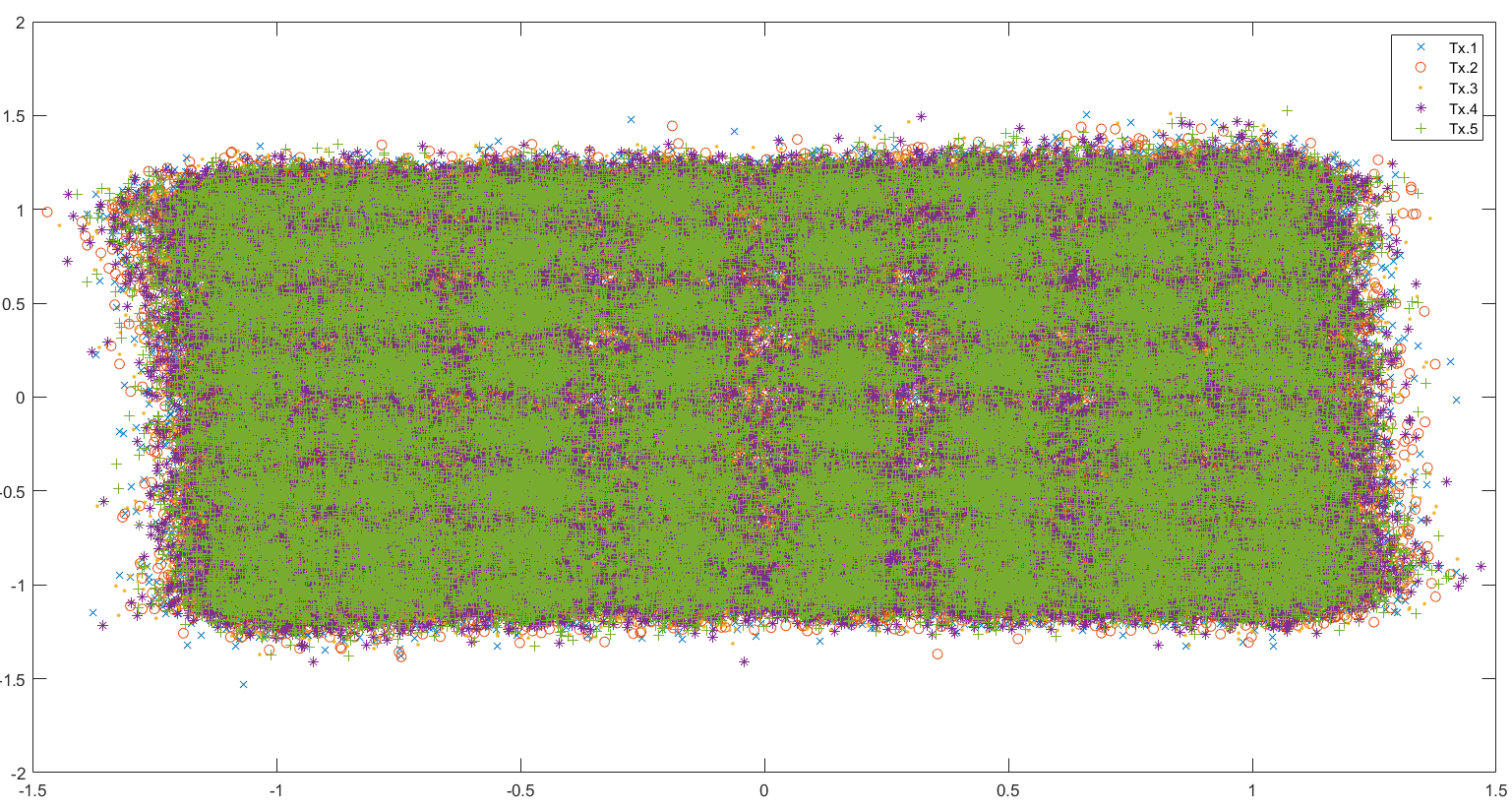}
        \caption{The data detected by the receiver from the noisiest (10.7dBm noise) channel. All 5 iterations of transmission are depicted together.}
        \label{fig:chap_data_10_7}
    \end{minipage}
    \end{figure}

    The dataset consists of a launch power sweep of 80~km optic fibre transmission of coherent 80~GBd dual polarization~(DP)-64QAM with a gross data rate of 960Gb/s. 
    15\% overhead for FEC and 3.47\% overhead for pilots and training sequences have been used, leading to a net bit rate of 800Gb/s. 
    The experimental setup to capture this real-world database is shown in \cref{fig:Setup}. 
    A more detailed explanation of the data and setup can be found in~\cite{stereo_paper_entropy}.
    The received raw signals have preprocessed by the receiver to create the dataset upon which clustering could be performed as a final decoding step. 
    The signals were also normalized to fit the initial transmission values (alphabet). 
    The data consists of 4 sets with different launch powers, corresponding to different noise levels during transmission: $2.7$dBm, $6.6$dBm, $8.6$dBm, and $10.7$dBm. 
    The average launch power in Watts (W) can be calculated as follows:
    \begin{align*}
        P_{(\mathrm{W})} = 1\mathrm{W} \cdot 10^{P_{(\mathrm{dBm})}/10} /1000 = 10^{(P_{(\mathrm{dBm})}-30)/10}\mathrm{W}.
    \end{align*}
    \Cref{fig:chap_data_2_7} shows the received data (all 5 instances of transmission) for the dataset with the least noise (2.7dBm), and \cref{fig:chap_data_10_7} shows the received data (all 5 instances of transmission) for the dataset with most noise (10.7dBm).

    For the generation of the 16-QAM data, the received signal is modelled as (amplitude damping is ignored):
    \begin{align}
        \hat{\mathbf{s}}~\coloneqq~e^{\mathrm{i}(\varphi_\mathrm{b}+\Phi)}\cdot~\mathbf{s}~+~\mathbf{N}
        \label{eq:noisy_signal}
        \;.
    \end{align}
    Here $\mathbf{s}$ is the transmitted signal, $\Phi$ is a random phase acquired during transmission which is distributed according to a normal distribution with zero mean and variance $\sigma_\Phi$~\cite{kpb2018-QuantumKerrModel}, $\mathbf{N}$ is additive zero mean Gaussian noise of variance $\sigma_\mathbf{N}$ (AWGN)~\cite{ghozlanKramer2010}, and $\varphi_\mathrm{b}$ is an unknown induced by birefringence that is assumed to be constant over some period of time.

\section{Methodology}
\label{sec:methods}

\newcommand{\eps}{\epsilon}

\subsection{Data Embedding Procedure} \label{subsec:data_embedding_procedure}

    Classical data needs to be converted into quantum states for processing in a quantum computer due to the poor coherence times and very limited number of qubits in current NISQ (Noisy Intermediate Scale Quantum) devices.
    One of the most popular methods of data encoding is angle embedding since it needs only $\mathcal{O}(1)$ operations regardless of how many data values need to be encoded.  
    Distance estimation between two data points $(x_1,y_1)$ and $(x_2,y_2)$ using angle embedding consists of preparing quantum states through a unitary operation with the data point encoded in it.
    The two dimensional data vectors are normalised and transformed as~\cite{stephen2021}
    \begin{align}
        x'_i &= \frac{\pi}{2}\left( \tfrac{{x}_i}{\sqrt{x_i^2 + y_i^2}} + 1 \right) 
        &&& 
        y'_i &= \frac{\pi}{2}\left( \tfrac{{y}_i}{\sqrt{x_i^2 + y_i^2}} + 1 \right)\;. \label{eq:x_y_angle_emb}
    \end{align}
    % where $ \{\Hat{x},\Hat{y}\}_i = \frac{\{x,y\}_i}{\sqrt{x_i^2 + y_i^2}}$.
    This mapping enables us to encode the data points as
    \begin{align}
        \ket\psi &= U(x_1', y_1')\ket{0} & \text{and} & & \ket\phi &= U(x_2', y_2')\ket{0}\;,
    \end{align}
    where $U$ is the unitary
    \begin{align}
        U(\theta, \gamma) \coloneqq 
            \left(\begin{tabular}{cc}
                $\cos\tfrac{\theta}{2}$ & $-\sin\tfrac{\theta}{2}$ \\
                $e^{i\gamma} \sin\tfrac{\theta}{2}$ & $e^{i\gamma}\cos\tfrac{\theta}{2}$
            \end{tabular}\right)\;. \label{eq:unitary}
    \end{align}
    
    However, one can see that this form of embedding is not injective. If two points in the plane lie on the same radial line, \cref{eq:x_y_angle_emb} will yield identical values of $x'_i$ and $y'_i$. This is not ideal for k-means clustering since it is quite possible to have several points with the same phase. To counter this problem, one can calculate and store the amplitude of each vector, and then use that to estimate the distance.
    
    Instead, we preprocess the data to achieve an injective data embedding scheme.
    We do so by using the following a different transformation for the 2-dimensional data vector:
    \begin{align}
        x''_i &= \frac{\pi}{2}\left( {\Bar{x}_i} + 1 \right) 
        &&& 
        y''_i &= \frac{\pi}{2}\left( {\Bar{y}_i} + 1 \right)\;,
        \label{eq:x_y_our_emb}
    \end{align}
    where $\{\Bar{x},\Bar{y}\}_i = \tfrac{\{x,y\}_i}{r_{max}}$ and $r_{\max}$ $\coloneqq \max_i \left\{ \sqrt{x_i^2 + y_i^2}\right\}$.
    The two transformed vectors are then still encoded in the same way (angle embedding) as
    \begin{align}
        \ket\psi &= U(x_1'', y_1'')\ket{0} 
        & & & 
        \ket\phi &= U(x_2'', y_2'')\ket{0}\;.
    \end{align}
    A distinct advantage of an injective embedding is that the output of the quantum circuit is eligible to be used directly for classifying the point, avoiding further post-processing steps. 
    In~\cite{stereo_paper_entropy}, we use the inverse stereographic projection to calculate the parameters of the embedding.

\subsection{Quantum circuit for overlap estimation}

    \begin{figure}
    
        \tikzcdset{row sep/normal = {1cm,between origins}}
        \[
        \begin{adjustbox}{max width = 0.5\textwidth}
        \begin{quantikz}[column sep = .16cm, align equals at = 2]
        \ket{0}\ &\qw & \gate{H} & \ctrl{1} & \gate{H} \gategroup[steps=2,style={dashed}]{} & \meterD{Z} &\cw
        \\
        % \ket{ψ}\ 
        \ket{0}\ & \gate{U(x_1'',y_1'', 0)} & \qw & \targX{ } & \qw & \qw 
        \\
         \ket{0}\ &  \gate{U(x_2'',y_2'', 0)} & \qw & \swap{-2} & \qw & \qw
        \end{quantikz}
        =
        \begin{quantikz}[column sep = .16cm, align equals at = 2]
        \ket{+}\ & \ctrl{1} & \meterD{X} &\cw
        \\
        \ket{\psi} & \targX{ } & \qw & \qw
        \\
        \ket{\phi} & \swap{-2} & \qw & \qw
        \end{quantikz}
          \end{adjustbox}
        \]
     \caption{Quantum circuit of the swap Test}
    \label{fig:swap_test_circuit}
    \end{figure}
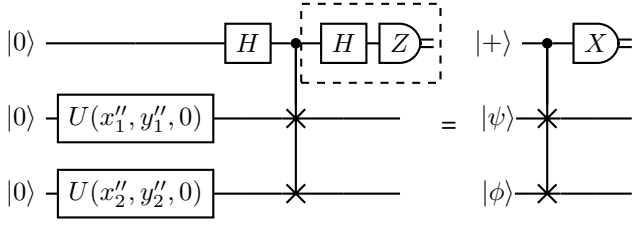

    Once the states have been prepared by encoding the classical data, they are processed through the quantum circuit shown in~\cref{fig:swap_test_circuit} to yield information about the overlap between the states. 
    This circuit corresponds to the swap test.
    The swap test is a method commonly used in quantum computing for learning overlap estimation between two quantum states, first developed in~\cite{BCWW01}.
    The task of finding the overlap is accomplished by measuring the output of the ancilla qubit of the swap test circuit many times. 
    The number of times this is repeated is known as the number of shots. 
    Since we only measure the ancilla, the other two qubits are theoretically available for future reuse; in other words, the swap test performs a non-destructive measurement. 
    We point out that this is wasteful in a hybrid quantum-classical system since quantum states are not stored or reused. 
    A better option, as mentioned in~\cite{stereo_paper_entropy}, is the Bell State measurement test, which performs a destructive measurement - saving not only a quantum gate but also a qubit. 

\subsection{Swap Test Probabilities and Distance Loss Function}

    In this section, we prove that the output of \cref{fig:swap_test_circuit} can be used for distance estimation, and calculate the end-to-end result of our procedure.
    Given two input states $\ket\psi$ and $\ket\phi$, independent of the form of embedding, the swap test yields a Bernoulli random variable $M$ with values in $\{0,1\}$ defined via:
    \begin{align}
     \mathbbm{P} (M=m) &= \frac{1}{2}(1 + (-1)^m |\langle\psi|\phi\rangle|^2) \label{eq:prob_0_1}
    \end{align}
     and variance
    \[\mathrm{Var}(M) = \mathbbm{P}(M=0) \cdot \mathbbm{P}(M=1) = \frac{1-|\langle\psi | \phi\rangle|^4}{4}. \]
    We use the unbiased estimator:
    \begin{align}
        \mathbbm{P} (M=1) \approx \frac{1}{n}\sum_{j=1}^nm_{j}, \label{eq:estimator}
    \end{align}
    where $m_{1},\ldots,m_{n}\in\{0,1\}$ are the measurement results obtained from the repeated measurement of the ancilla qubit of the swap test circuit. 
    
\begin{figure}
    \begin{minipage}{0.24\textwidth}
    \includegraphics[width=1.0\textwidth]{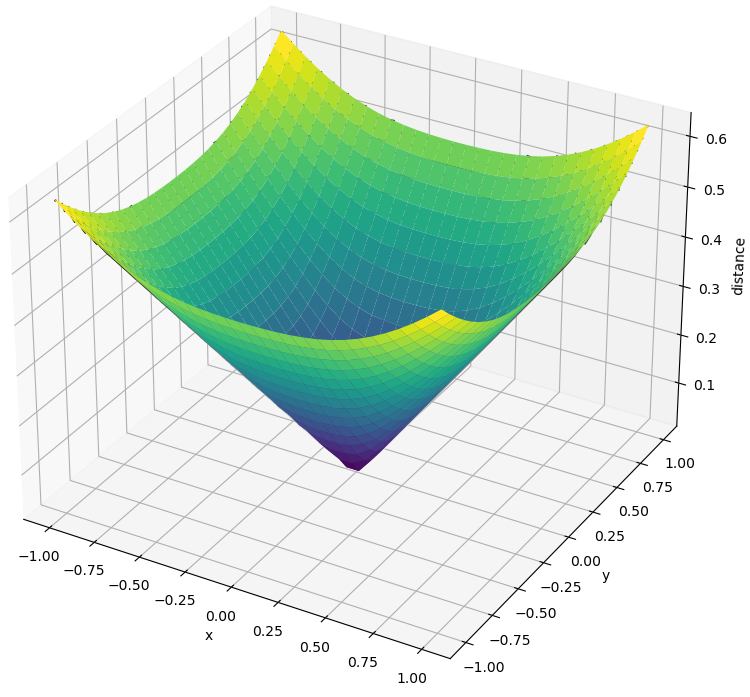}
    \caption{Standard angle embedding distance loss function between $(x,y)$ and $(0,0)$}\label{fig:ang_emb_dist}
    \end{minipage}\hfill
    \begin{minipage}{0.24\textwidth}
    \includegraphics[width=1.0\textwidth]{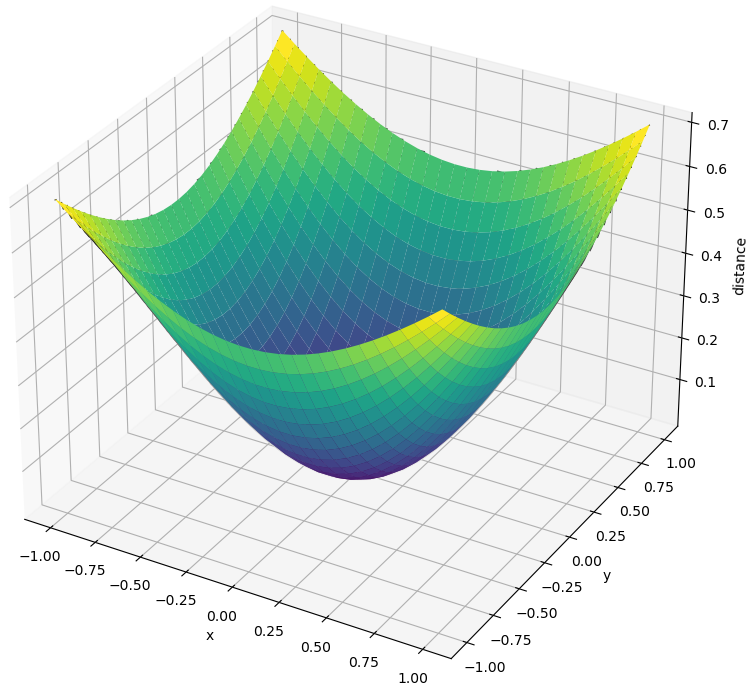}
    \caption{Our version of the Angle embedding loss function between $(x,y)$ and $(0,0)$ with $r_{max}=1$}\label{fig:ang_new_emb_dist}
    \end{minipage}
    % \end{figure}
    
    % \begin{figure}%[!htb]
    \begin{minipage}{0.24\textwidth}
    \includegraphics[width=1.0\textwidth]{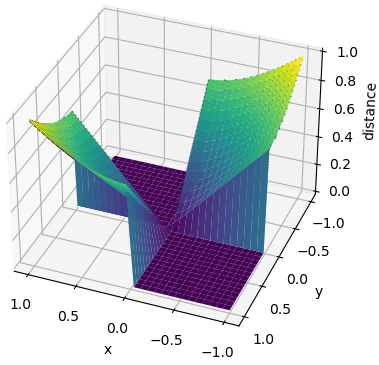}
    \caption{Standard angle embedding distance loss function for 2 points on the x-axis}\label{fig:ex_1_why_better}
    \end{minipage}\hfill
    \begin{minipage}{0.24\textwidth}
    \includegraphics[width=1.0\textwidth]{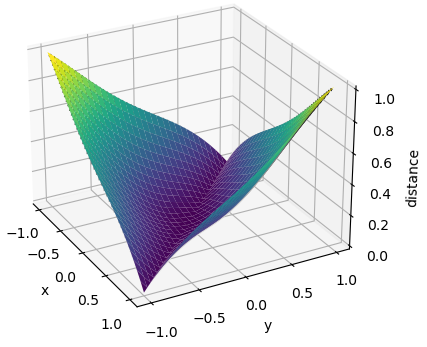}
    \caption{Our version of the Angle embedding loss function for 2 points on the x-axis with $r_{max}=1$}\label{fig:ex_2_why_better}
    \end{minipage}
    \end{figure}

    After assuming the form of embedding described in \cref{subsec:data_embedding_procedure}, one can calculate the expression for the projection between the two qubits as follows:
    \begin{align}
    \bra{\phi}\ket{\psi} 
    &= 
    \cos\left( \frac{\pi}{4}(\Bar{x}_1+1) \right) 
    \cos\left( \frac{\pi}{4}(\Bar{x}_2+1) \right) 
    \nonumber\\
    & + e^{i\tfrac{\pi}{2}(\Bar{y}_1 - \Bar{y}_2)}
    \sin\left( \frac{\pi}{4}(\Bar{x}_1+1) \right) 
    \sin\left( \frac{\pi}{4}(\Bar{x}_2+1) \right)\;.
    \label{eq:overlap}
    \end{align}
    Using this, we get our final distance estimate or `distance loss function':
    \begin{align}
    \mathbbm{P}(1) &
    % = \frac{1}{2}\left(1 - |\langle \psi | \phi \rangle |^2\right) 
    = \frac{1}{4}\left(1 - \cos\left( \frac{\pi}{2}\Bar{x} \right)\cos\left( \frac{\pi}{2}\Bar{y} \right)\right) \label{eq:DLF}
    \end{align} 
    One can see through \cref{eq:DLF} that the hybrid quantum-classical implementation of this algorithm provides a different loss function for the distance between $(x_1,y_1)$ and $(x_2,y_2)$ than the Euclidean distance. 
    The overlap $\bra{\phi}\ket{\psi}$ is estimated from the probabilities of the swap test and the expression of the estimator (\cref{eq:prob_0_1,eq:estimator}).
    As we can see from \cref{eq:DLF}, the swap test using both standard angle embedding and our form of it does not enable calculation of the true Euclidean distance.
    It uses, in fact, a completely different distance estimate. 
    It is illustrative to pick the point $(0,0)$ (origin) and see how the distance loss function with other points $(x,y)$ varies - see \cref{fig:ang_emb_dist,fig:ang_new_emb_dist}. 
    One can see that with our method of rescaling, the loss function is smoother at origin.
    The advantage of our method can be brought into sharper focus with another example - distance estimation between $(x,0)$ and $(y,0)$ i.e., between 2 points on the x-axis. Traditional angle embedding yields the loss function depicted in \cref{fig:ex_1_why_better} while our version yields \cref{fig:ex_2_why_better}.   
    
    The performance of the swap test can also be increased by using more elaborate testing strategies~\cite{fanizzaRosati-beyond-swap}. Performance bounds are made visible in~\cite[Fig.~2]{fanizzaRosati-beyond-swap}. However, from the NISQ perspective, the joint measurements over many copies of the quantum states representing the data points contradicts the idea of a simplified system design and limiting the number of qubits and gates.

\section{Performance and Results}
\label{sec:results}

\subsection{Characterisation using 16-QAM data}

    The accuracy of the quantum $k$-means code is first reviewed through experiments performed on \textit{generated $16$-QAM noisy data}. Figures \ref{fig:16-QAM for phase noise 0.05, Gaussian (radial) noise=0.01, Overall Rotation = 0} and \ref{fig:16-QAM for phase noise 0.2, Gaussian (radial) noise=0.04, Overall Rotation = 3pi/10} are representative of the kind of 16-QAM datasets upon which the algorithms were performed.

    \begin{figure}
    \begin{minipage}{0.24\textwidth}
    
        \includegraphics[width=\linewidth]{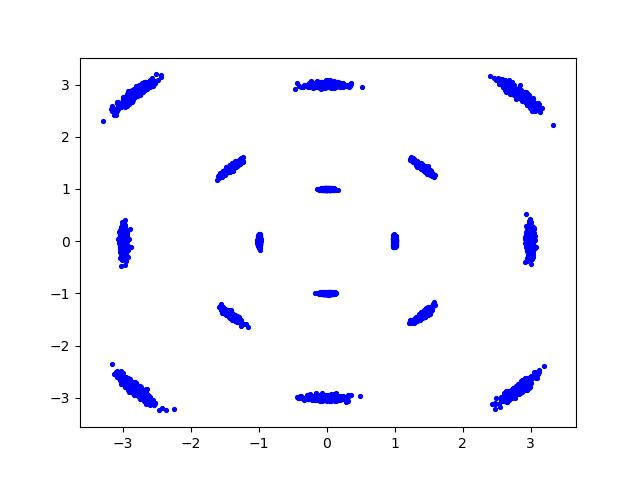}
        \caption{ Parameters are\\$\sigma_\Phi=0.05$, $\sigma_\mathbf{N}=0.01$, $\varphi_\mathrm{b}=0$.}\label{fig:16-QAM for phase noise 0.05, Gaussian (radial) noise=0.01, Overall Rotation = 0}
     
    \end{minipage}
    \begin{minipage}{0.24\textwidth}
    
        \includegraphics[width=\linewidth]{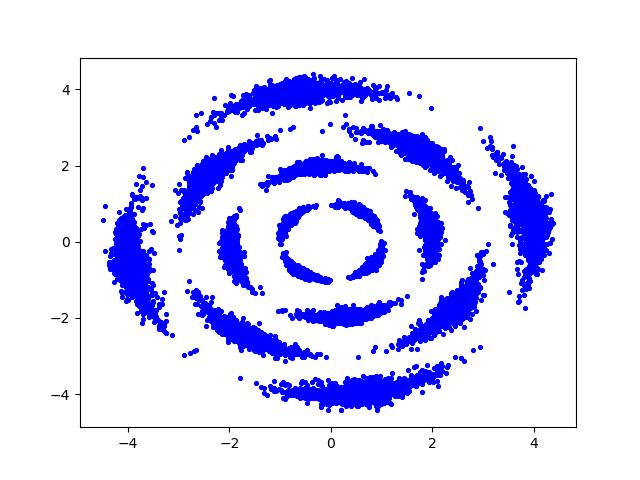}
        \caption{  Parameters are\\$\sigma_\Phi=0.2$, $\sigma_\mathbf{N}=0.04$, $\varphi_\mathrm{b}=\frac{3\pi}{10}$. }\label{fig:16-QAM for phase noise 0.2, Gaussian (radial) noise=0.04, Overall Rotation = 3pi/10}
    
    \end{minipage}
    \end{figure}

    The first test shows how the number of shots used for the swap-test affects accuracy (see \cref{fig:shots-vs-acc}).
    Next, we demonstrate the effect of the fixed phase noise of the data on the accuracy (see \cref{fig:-and-rot-vs-acc}).
    Lastly, an accuracy heat-map by varying the noise levels is produced (shown in \cref{fig:acc-surf}).
    The confusion matrix of the experiment is shown in \cref{fig:clustering-cartisian}. 
    
    \begin{figure}%[!htb]
    \centering
    {% This file was created with tikzplotlib v0.9.12.
% % axis style, ticks, etc
% \pgfplotsset{every axis/.append style={
%                     axis x line=middle,    % put the x axis in the middle
%                     axis y line=middle,    % put the y axis in the middle
%                     axis line style={<->}, % arrows on the axis
%                     xlabel={$x$},          % default put x on x-axis
%                     ylabel={$y$},          % default put y on y-axis
%                     label style={font=\tiny},
%                     tick label style={font=\tiny}  
%                     }}

\small

\begin{tikzpicture}
\begin{axis}[
    scale=1,
    tick align=outside,
    tick pos=left,
    x grid style={white!69.0196078431373!black},
    xmajorgrids,
    ymajorgrids,
    legend style=
    {at={(0.92,0.02)},anchor=south east, 
    nodes={scale=0.75, transform shape}},
    height=5.6cm,
    width=8.7cm,
    xlabel={Number of Swap-Test Shots},
    ylabel={Clustering Accuracy},
    xtick style={color=black},
    y grid style={white!69.0196078431373!black},
    ytick style={color=black},
    xmode=log,
    log basis x={2},
    legend columns=2
]
\addplot [very thick, red!90!black, smooth]
table {%
8 0.11405473602484473
16 0.2098906951751379
32 0.7129939850227321
64 0.9603565365025467
128 0.9899919871794872
256 1.0
512 1.0
1024 1.0
2048 1.0
};
\addlegendentry{($\sigma_\Phi$,$\sigma_N$)=(0.1,0.1)}

\addplot [very thick, blue!90!white, smooth]
table[col sep=comma] {%
8, 0.1135684327600733
16, 0.2237353639008051
32, 0.6627862146704058
64, 0.9458285797153632
128, 0.9714043399264208
256, 0.9905801435406698
512, 0.9813854895104895
1024, 0.9875218781185207
2048, 0.9975961538461539
};
\addlegendentry{($\sigma_\Phi$,$\sigma_N$)=(0.2,0.2)}

\addplot [very thick, green!50!black, smooth]
table[col sep=comma] {%
8, 0.13361742424242423
16, 0.25948307554186967
32, 0.4734851881706908
64, 0.783860403584468
128, 0.8245779166009429
256, 0.8772599123639366
512, 0.839001421685528
1024, 0.8951704026472249
2048, 0.8644003603582409
};
\addlegendentry{($\sigma_\Phi$,$\sigma_N$)=(0.3,0.3)}

\addplot [very thick, purple!70!white, smooth]
table[col sep=comma] {%
8, 0.10556519503887925
16, 0.1904094484221352
32, 0.31873321837481616
64, 0.5268633268706191
128, 0.5011691388858781
256, 0.5495917776305446
512, 0.5835766847634598
1024, 0.5494454082340238
2048, 0.6267316594661836
};
\addlegendentry{($\sigma_\Phi$,$\sigma_N$)=(0.4,0.4)}

\end{axis}

\end{tikzpicture}}
    \caption{Number of shots vs. Accuracy}\label{fig:shots-vs-acc}
    \par\vspace{.5em}
    {% This file was created with tikzplotlib v0.9.12.
\small
\begin{tikzpicture}

\definecolor{color0}{rgb}{0.12156862745098,0.466666666666667,0.705882352941177}
\definecolor{color1}{rgb}{1,0.498039215686275,0.0549019607843137}

\begin{axis}[
tick align=outside,
tick pos=left,
x grid style={white!69.0196078431373!black},
legend style={
    at={(0.98, 0.4)},
    anchor=south east,
    nodes={scale=0.75, transform shape}
},
xmajorgrids,
ymajorgrids,
height=5.7cm,
width=8.7cm,
xlabel={Phase},
ylabel={Clustering Accuracy},
xtick style={color=black},
y grid style={white!69.0196078431373!black},
ytick style={color=black},
legend columns=2
]

\addplot [very thick, red!90!black, smooth]
table[col sep=comma] {%
-0.39269908169872414, 1
-0.3365992128846207, 1
-0.28049934407051724, 1
-0.2243994752564138, 1
-0.16829960644231035, 1
-0.1121997376282069, 1
-0.05609986881410345, 1
0.0, 1
0.05609986881410345, 1
0.1121997376282069, 1
0.16829960644231035, 1
0.2243994752564138, 1
0.28049934407051724, 1
0.3365992128846207, 1
0.39269908169872414, 1
};
\addlegendentry{($\sigma_\Phi$,$\sigma_N$)=(0.1,0.1)}

\addplot [very thick, blue!90!white, smooth]
table[col sep=comma] {%
-0.39269908169872414, 0.9967105263157895
-0.3365992128846207, 0.9894610507246377
-0.28049934407051724, 0.9946590909090909
-0.2243994752564138, 0.9866510025062656
-0.16829960644231035, 0.9934826203208557
-0.1121997376282069, 0.9904605263157895
-0.05609986881410345, 0.9885993083003952
0.0, 0.9873949579831933
0.05609986881410345, 0.9945549242424243
0.1121997376282069, 0.9944196428571429
0.16829960644231035, 0.99375
0.2243994752564138, 0.9822176179084073
0.28049934407051724, 0.9925881410256411
0.3365992128846207, 0.992378762541806
0.39269908169872414, 0.9979166666666667
};
\addlegendentry{($\sigma_\Phi$,$\sigma_N$)=(0.2,0.2)}

\addplot [very thick, green!50!black, smooth]
table[col sep=comma] {%
-0.39269908169872414, 0.8252354827985326
-0.3365992128846207, 0.869898043918408
-0.28049934407051724, 0.869266984371053
-0.2243994752564138, 0.912291816979317
-0.16829960644231035, 0.8749662663725164
-0.1121997376282069, 0.8716870139664257
-0.05609986881410345, 0.9315222799855484
0.0, 0.8757507292889739
0.05609986881410345, 0.8985535644910645
0.1121997376282069, 0.913922305520307
0.16829960644231035, 0.8659821593013097
0.2243994752564138, 0.8275330803751519
0.28049934407051724, 0.8413041644460629
0.3365992128846207, 0.835767888116242
0.39269908169872414, 0.8738813111745541
};
\addlegendentry{($\sigma_\Phi$,$\sigma_N$)=(0.3,0.3)}

\addplot [very thick, color0, smooth]
table[col sep=comma] {%
-0.39269908169872414, 0.5393447508452711
-0.3365992128846207, 0.4925400940662841
-0.28049934407051724, 0.536203621318158
-0.2243994752564138, 0.529366091293273
-0.16829960644231035, 0.591085126588836
-0.1121997376282069, 0.5504280369411507
-0.05609986881410345, 0.5754281922718016
0.0, 0.616017327372826
0.05609986881410345, 0.5979704714609311
0.1121997376282069, 0.6069925379572119
0.16829960644231035, 0.5958906879400468
0.2243994752564138, 0.574336344512975
0.28049934407051724, 0.5273918158270835
0.3365992128846207, 0.5157481614790684
0.39269908169872414, 0.47978966819140434
};
\addlegendentry{($\sigma_\Phi$,$\sigma_N$)=(0.5,0.5)}
\end{axis}

\end{tikzpicture}}
    \caption{Mean Phase Noise vs. Accuracy}\label{fig:-and-rot-vs-acc}
    \end{figure}
    
    \begin{figure}
    \centering
    {\small
\begin{tikzpicture}
\begin{axis}[
    scale = 0.9,
    view = {120}{20},
    tick align=outside,
    tick pos=left,
    x grid style={white!69.0196078431373!black},
    xmajorgrids,
    ymajorgrids,
    zmin=0, zmax=1.1,
    xlabel={$\sigma_N$},%{Amplitude Noise ($\sigma_N$)},
    ylabel={$\sigma_\Phi$},%{Angular Noise $\sigma_\Phi$},
    zlabel={Algorithm Accuracy}
    ]
    \addplot3 [
    surf,
    shader=faceted interp,
    mesh/cols=15,
    draw=gray] 
    table {plots/accuracy_data.txt};
\end{axis}
\end{tikzpicture}}
    \caption{Heat-map plot of the accuracy with varying noise parameters.}\label{fig:acc-surf}
    \par\vspace{1em}
    \includegraphics[width=.4\textwidth]{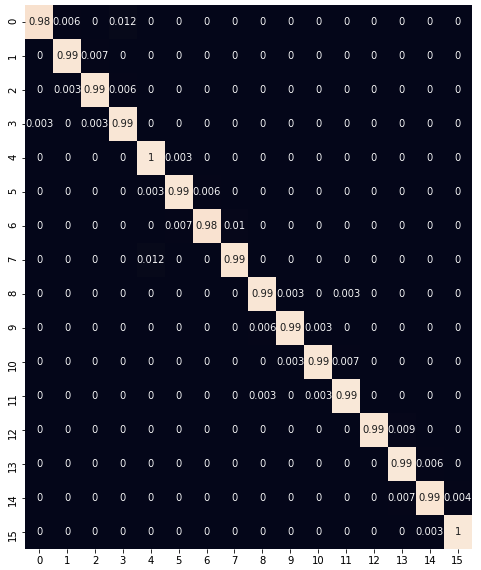} 
    \caption{The accuracy of performing the quantum clustering algorithm with angle embedding in simulation using 5 distance estimates per circuit.}\label{fig:clustering-cartisian}
    \end{figure}

\subsection{Testing using optical fibre real data}

% \vspace{0.1cm}
    \begin{table}
    \centering
    \begin{tabular}{|c|c|c|c|c|c|}
        \hline
         \multicolumn{2}{|c|}{Total No. of Points} & \multicolumn{2}{|c|}{Accuracy (\%)} &  \multicolumn{2}{|c|}{Max. iterations} \\
         \hline
         Quantum  & Classical & Quantum  & Classical    &  Quantum & Classical \\
        \hline
        320   & 320   & 79.5 & 88.5  & 5  & 5\\ 
        \hline
        640   & 640  & 83.4  & 88.5  &  5 & 5\\
        \hline
        1280   & 1280  & 83.5  & 87.3  &  5 & 5\\
        \hline 
    \end{tabular}
    \vspace{.5em}
    \caption{Experiments for 2.7 dBm}
    \label{tab:2_7_dB_summary_table}
    \begin{tabular}{|c|c|c|c|c|c|}
        \hline
         \multicolumn{2}{|c|}{Total No. of Points} & \multicolumn{2}{|c|}{Accuracy (\%)} &  \multicolumn{2}{|c|}{Max. iterations} \\
         \hline
         Quantum  & Classical & Quantum  & Classical     &  Quantum & Classical \\
        \hline
        320   & 320   & 78.0 & 85.2  &  5  & 5\\ 
        \hline
        640   & 640  & 82.0  & 87.7   &  5 & 5\\
        \hline
        1280   & 1280  & 82.4  & 87.7  &  5 & 5\\
        \hline
    \end{tabular}
    \vspace{.5em}
    \caption{Experiments for 6.6 dBm}
    \label{tab:6_6_dB_summary_table}
    \begin{tabular}{|c|c|c|c|c|c|}
        \hline
         \multicolumn{2}{|c|}{Total No. of Points} & \multicolumn{2}{|c|}{Accuracy (\%)} &  \multicolumn{2}{|c|}{Max. iterations} \\
         \hline
         Quantum  & Classical & Quantum  & Classical   &  Quantum & Classical \\
        \hline
        320   & 320   & 79.1 & 87.4    & 5  & 5\\ 
        \hline
        640   & 640  & 80.9  & 87.1    & 5 & 5\\
        \hline
        1280   & 1280  & 83.2  & 87.6   & 5 & 5\\
        \hline
    \end{tabular}
    \vspace{.5em}
    \caption{Experiments for 8.6 dBm}
    \label{tab:8_6_dB_summary_table}
    \begin{tabular}{|c|c|c|c|c|c|}
        \hline
         \multicolumn{2}{|c|}{Total No. of Points} & \multicolumn{2}{|c|}{Accuracy (\%)} & \multicolumn{2}{|c|}{Max. iterations} \\
         \hline
         Quantum  & Classical & Quantum  & Classical  &  Quantum & Classical \\
        \hline
        320   & 320   & 72.6 & 80.2    & 5  & 5\\ 
        \hline
        640   & 640  & 74.1  & 82.5  & 5 & 5\\
        \hline
        1280   & 1280  & 77.7  & 82.3 & 5 & 5\\
        \hline
    \end{tabular}
    \vspace{0.5em}
    \caption{Experiments for 10.7 dBm}
    \label{tab:10_7_dB_summary_table}
    \end{table}

    As mentioned before, the experimentally collected real-world data is 64-QAM data transmitted via optic fibre, and there are 4 sets of data with different noise levels during transmission: $2.7$dBm, $6.6$dBm, $8.6$dBm, and $10.7$dBm. 
    % We tested the developed hybrid quantum-classical and classical $k$-means clustering code on this data while varying various parameters.
    \Cref{tab:2_7_dB_summary_table,tab:6_6_dB_summary_table,tab:8_6_dB_summary_table,tab:10_7_dB_summary_table} summarize the accuracy, number of points and maximum number of allowed algorithm iterations for datasets with channel noises of $2.7$, $6.6$, $8.6$ and $10.7$dBm respectively. 
    The number of shots for overlap estimation using the quantum circuit was kept constant at 50 shots.
    
    As seen in \cref{tab:2_7_dB_summary_table,tab:6_6_dB_summary_table,tab:8_6_dB_summary_table,tab:10_7_dB_summary_table}, the hybrid quantum-classical implementation of quantum $k$-means clustering using angle embedding performs noticeably worse than classical $k$-means in terms of accuracy.
    This accuracy discrepancy stems from \emph{the process of data embedding}. 
    The use of angle embedding caused the distance estimated using the loss function from \cref{eq:DLF} to become more susceptible to noise in the optic fibre cable.
    This is further discussed in the next section.
    
    The time taken for execution of the hybrid quantum-classical and classical $k$-means clustering algorithms on the real data differ greatly due to the high computational time of simulating quantum devices in a classical computer. 
    This implementation can be applied into programming a real quantum chip, which could improve the time performance.
    However, currently quantum gate delays are $\sim 1000\times$ classical gate delays.
    An estimation of the processing time it would take to cluster a 16-QAM dataset with 5000 2D data points using angle encoding follows.
    Gate times for the 7-qubit IBMQ Casablanca are in the range of 305 - 760 ns, the average gate time being 443 ns.
    Using the IBM Quantum Platform, which uses the available basis gates for implementation, we obtain that for angle embedding the circuit depth is 22.
    Hence we estimate the time taken for one shot of the swap test to be from 6710 to 16720 ns, and 9746 ns on average. 
    To compute 1 iteration with 16 centroids, 5000 datapoints and 50 shots per distance estimation, the number of swap test shots needed is $16 \cdot 5000 \cdot 50 = 4 \cdot 10^6$. 
    Ignoring the pre-processing steps, the QPU time will therefore be $26.84$ to $66.88$ seconds, $38.984$ seconds on average - much slower than current classical computers.
    As the quantum hardware technology node advances and gate times reduce (faster circuits, more stable qubits, less quantum error correction), the quantum algorithm will become more competitive time-wise.

\section{Discussion and Conclusion}
\label{sec:conclusions}

    The core idea for the paper was to analyse the strengths and weaknesses of the quantum $k$-means clustering proposed by Lloyd et al.\ in the NISQ context to cluster QAM data. 
    The proposed algorithm in~\cite{lloyd2013quantum} assumes a quantum random access memory for state preparation, which allows them to access the data in quantum parallel.
    Since at this time there are no commercially available practical realisations of such a quantum RAM, we decided to use a hybrid quantum-classical approach, where only the distance estimation part of the algorithm is done using a quantum computer. 
    The rest of the algorithm is executed on a classical computer. 
    Naturally, this required us to embed classical data into quantum states - this introduces the classical data loading problem, leading us to question if there really is any quantum advantage.
    For the preparation of quantum states, we decided to use angle embedding since it has a simple $\mathcal{O}$(1) implementation using current basis gates. 
    Using the unitary as defined in \cref{eq:unitary} leads effectively to a $M$-way classification using a cosine kernel~\cite{BPRP20, schuld21supervised}.
    This makes the expectation of quantum advantage seem more suspect since a cosine kernel can be easily computed classically. 
    We also show that when using the defined unitary (a product of rotation gates with a global phase), one does not produce the Euclidean distance but (as one would expect from a cosine kernel) a trigonometric `loss function'.
    We predict the loss function, and we see that this `distance loss function' is less steep than a paraboloid - this led us to believe that the performance would be adversely affected. 
    This prediction is indeed supported by the simulation.
    To stress the point further: \emph{we predict and show that the hybrid quantum-classical k-means algorithm has an inferior performance to classical k-means simply due to the process of angle embedding}. 
    This is a theoretically predicted loss in performance; even when the simulations are performed without quantum noise, the algorithm's performance is expected to be deficient. 
    With the addition of quantum noise, one would expect the accuracy of the hybrid quantum-classical algorithm to suffer even further.
    However, in spite of this worse loss function, we see that the accuracy is surprisingly high.
    Despite a completely different distance estimate, our accuracy was on par in the case of 16-QAM and slightly worse for 64-QAM. 
    It seems that as the non-linear noise and the number of clusters increases, the performance of the angle embedding loss function decreases.

    Our approach opens the door for other distance loss functions which could yield even better accuracy than the classical algorithm. 
    This approach is similar to that of constructing `Quantum Kernels' for SVM and other kernel based learning methods~\cite{IBM_QSVM_2019, LAT21}. 
    We can construct `quantum distance loss functions' which can then be used for clustering of data. 
    These functions can also be used in applications such as spectral clustering or nearest mean classification. 
    Such other kinds of data embedding, where the unavoidable step of data embedding can in fact be used to one's advantage, is explored in~\cite{stereo_paper_entropy} in particular, it discusses the very promising stereographic embedding.
    
    Another important direction of future work is to benchmark the performance of the algorithms on standard clustering datasets. 
    An investigation to be carried out is to find use-cases better tailored to the hybrid implementation.

\section*{Acknowledgement}

    This work was funded by the TUM-Huawei Joint Lab on Algorithms for Short Transmission Reach Optics (ASTRO). This project has received funding from the DFG Emmy-Noether program under grant number NO 1129/2-1 (JN) and by the Federal Ministry of Education and Research of Germany in the programme of ”Souveran.\ Digital.\ Vernetzt.”. Joint project 6G-life, project identification number: 16KISK002, and of the Munich Center for Quantum Science and Technology (MCQST). We would also like to acknowledge fruitful discussions with Stephen DiAdamo and Fahreddin Akalin during the initial stages of the project.

% \bibliography{biblio.bib}
\bibliographystyle{IEEEtran}
\bibliography{bibliography}

\end{document}